\PassOptionsToPackage{unicode}{hyperref}
\PassOptionsToPackage{hyphens}{url}
\PassOptionsToPackage{dvipsnames,svgnames,x11names}{xcolor}
\documentclass[
]{article}
\usepackage{amsmath,amssymb}
\usepackage{lmodern}
\usepackage{iftex}
\ifPDFTeX
  \usepackage[T1]{fontenc}
  \usepackage[utf8]{inputenc}
  \usepackage{textcomp} 
\else 
  \usepackage{unicode-math}
  \defaultfontfeatures{Scale=MatchLowercase}
  \defaultfontfeatures[\rmfamily]{Ligatures=TeX,Scale=1}
\fi
\IfFileExists{upquote.sty}{\usepackage{upquote}}{}
\IfFileExists{microtype.sty}{
  \usepackage[]{microtype}
  \UseMicrotypeSet[protrusion]{basicmath} 
}{}
\makeatletter
\@ifundefined{KOMAClassName}{
  \IfFileExists{parskip.sty}{%
    \usepackage{parskip}
  }{
    \setlength{\parindent}{0pt}
    \setlength{\parskip}{6pt plus 2pt minus 1pt}}
}{
  \KOMAoptions{parskip=half}}
\makeatother
\usepackage{xcolor}
\usepackage{color}
\usepackage{fancyvrb}

\DefineVerbatimEnvironment{Highlighting}{Verbatim}{commandchars=\\\{\}}
\newenvironment{Shaded}{}{}

\newcommand{\CommentTok}[1]{\textcolor[rgb]{0.38,0.63,0.69}{\textit{#1}}}

\newcommand{\DecValTok}[1]{\textcolor[rgb]{0.25,0.63,0.44}{#1}}

\newcommand{\FloatTok}[1]{\textcolor[rgb]{0.25,0.63,0.44}{#1}}

\newcommand{\ImportTok}[1]{\textcolor[rgb]{0.00,0.50,0.00}{\textbf{#1}}}

\newcommand{\NormalTok}[1]{#1}
\newcommand{\OperatorTok}[1]{\textcolor[rgb]{0.40,0.40,0.40}{#1}}

\newcommand{\StringTok}[1]{\textcolor[rgb]{0.25,0.44,0.63}{#1}}

\usepackage{graphicx}
\makeatletter
\def\maxwidth{\ifdim\Gin@nat@width>\linewidth\linewidth\else\Gin@nat@width\fi}
\def\maxheight{\ifdim\Gin@nat@height>\textheight\textheight\else\Gin@nat@height\fi}
\makeatother
\setkeys{Gin}{width=\maxwidth,height=\maxheight,keepaspectratio}
\makeatletter
\def\fps@figure{htbp}
\makeatother
\NewDocumentCommand\citeproctext{}{}
\NewDocumentCommand\citeproc{mm}{%
  \begingroup\def\citeproctext{#2}\cite{#1}\endgroup}
\makeatletter
 \let\@cite@ofmt\@firstofone
 \def\@biblabel#1{}
 \def\@cite#1#2{{#1\if@tempswa , #2\fi}}
\makeatother
\newlength{\cslhangindent}
\newlength{\csllabelwidth}
\newenvironment{CSLReferences}[2] 
 {\begin{list}{}{%
  \setlength{\itemindent}{0pt}
  \setlength{\leftmargin}{0pt}
  \setlength{\parsep}{0pt}
  \ifodd #1
   \setlength{\leftmargin}{\cslhangindent}
   \setlength{\itemindent}{-1\cslhangindent}
  \fi
  \setlength{\itemsep}{#2\baselineskip}}}
 {\end{list}}
\usepackage{calc}

\ifLuaTeX
\usepackage[bidi=basic]{babel}
\else
\usepackage[bidi=default]{babel}
\fi
\babelprovide[main,import]{american}

\def\languageshorthands#1{}
\ifLuaTeX
  \usepackage{selnolig}  
\fi
\IfFileExists{bookmark.sty}{\usepackage{bookmark}}{\usepackage{hyperref}}
\IfFileExists{xurl.sty}{\usepackage{xurl}}{} 
\hypersetup{
  pdftitle={AstronomyCalc: A python toolkit for teaching Astronomical
Calculations and Data Analysis methods},
  pdfauthor={Sambit K. Giri},
  pdflang={en-US},
  colorlinks=true,
  linkcolor={Maroon},
  filecolor={Maroon},
  citecolor={Blue},
  urlcolor={Blue},
  pdfcreator={LaTeX via pandoc}}

\title{AstronomyCalc: A python toolkit for teaching Astronomical
Calculations and Data Analysis methods}

\definecolor{c53baa1}{RGB}{83,186,161}
\definecolor{c202826}{RGB}{32,40,38}

\usepackage[affil-it]{authblk}
\usepackage{orcidlink}
\author[1%
  ]{Sambit K. Giri%
    \,\orcidlink{0000-0002-2560-536X}\,%
    }

\affil[1]{Nordita, KTH Royal Institute of Technology and Stockholm
University, Hannes Alfvéns väg 12, SE-106 91 Stockholm, Sweden%
  }
\date{26 August 2024}

\begin{document}
\maketitle

\begin{center}
Report Number: NORDITA-2025-001
\end{center}

\section{Summary}\label{summary}

Understanding astrophysical and cosmological processes can be
challenging due to their complexity and the lack of simple, everyday
analogies. To address this, we present \texttt{AstronomyCalc}, a
user-friendly Python package designed to facilitate the learning of
these processes and help develop insights based on the variation theory
of learning (\citeproc{ref-ling2012variation}{Ling Lo, 2012};
\citeproc{ref-lo2011towards}{Lo \& Marton, 2011}).

\texttt{AstronomyCalc} enables students and educators to engage with key
astrophysical and cosmological calculations, such as solving the
Friedmann equations, which are fundamental to modeling the dynamics of
the universe. The package allows users to construct and explore various
cosmological models, including the de Sitter and Einstein-de Sitter
universes (see \citeproc{ref-ryden2017introduction}{Ryden, 2017} for
more examples), by adjusting key parameters such as matter density and
the Hubble constant. This interactive approach helps users intuitively
grasp how variations in these parameters affect properties like
expansion rates and cosmic time evolution.

Moreover, \texttt{AstronomyCalc} includes modules for generating
synthetic astronomical data or accessing publicly available datasets. In
its current version, users can generate synthetic Type Ia supernova
measurements of cosmological distances
(\citeproc{ref-vanderplas2012introduction}{VanderPlas et al., 2012}) or
utilize the publicly available Pantheon+ dataset
(\citeproc{ref-brout2022pantheonplus}{Brout et al., 2022}).
Additionally, the package supports the download and analysis of the
SPARC dataset, which contains galaxy rotation curves for 175 disk
galaxies (\citeproc{ref-lelli2016sparc}{Lelli et al., 2016}).

These datasets can be analyzed within the package to test cosmological
and astrophysical models, offering a hands-on experience that mirrors
the scientific research process in astronomy. \texttt{AstronomyCalc}
implements simplified versions of advanced data analysis algorithms,
such as the Metropolis-Hastings algorithm for Monte Carlo Markov chains
(\citeproc{ref-robert2004metropolis}{Robert \& Casella, 2004}), to
explain the fundamental workings of these methods. By integrating
theoretical concepts with observational data analysis,
\texttt{AstronomyCalc} not only aids in conceptual learning but also
provides insights into the empirical methods used in the field.

\section{Statement of Need}\label{statement-of-need}

The field of astronomy and cosmology requires a deep understanding of
complex processes that are often difficult to visualize or grasp through
traditional learning methods. \texttt{AstronomyCalc} addresses this
challenge by offering an interactive, user-friendly tool that bridges
the gap between theoretical knowledge and practical application.

Designed with the variation theory of learning in mind, this package
enables students and educators to experiment with and explore key
astrophysical and cosmological models in an intuitive manner. By varying
parameters and observing the resulting changes, users can develop a more
profound understanding of the underlying physical processes.

Furthermore, \texttt{AstronomyCalc} equips users with the tools needed
to analyze real or simulated astronomical data, thereby providing a
comprehensive learning experience that reflects the true nature of
scientific inquiry in astronomy. This makes \texttt{AstronomyCalc} an
invaluable resource for education in astronomy and cosmology, enhancing
both the depth and quality of learning in these fields.

\section{Usage Example}\label{usage-example}

To illustrate the functionality of \texttt{AstronomyCalc}, we will
estimate cosmological distances in different model universes.
Specifically, we will consider a dark energy-dominated universe (de
Sitter model), a matter-dominated universe (Einstein-de Sitter model),
and a universe based on the cosmology inferred from cosmic microwave
background radiation measurements by the Planck satellite.

\begin{Shaded}
\begin{Highlighting}[]
\ImportTok{import}\NormalTok{ numpy }\ImportTok{as}\NormalTok{ np }
\ImportTok{import}\NormalTok{ matplotlib.pyplot }\ImportTok{as}\NormalTok{ plt}
\ImportTok{import}\NormalTok{ AstronomyCalc }\ImportTok{as}\NormalTok{ astro}

\CommentTok{\# Redshift bins}
\NormalTok{zarr }\OperatorTok{=}\NormalTok{ np.logspace(}\OperatorTok{{-}}\DecValTok{2}\NormalTok{, }\DecValTok{1}\NormalTok{, }\DecValTok{100}\NormalTok{)}

\CommentTok{\# Einsten{-}de Sitter universe}
\NormalTok{cosmo\_dict }\OperatorTok{=}\NormalTok{ \{}\StringTok{\textquotesingle{}Om\textquotesingle{}}\NormalTok{: }\FloatTok{1.0}\NormalTok{, }\StringTok{\textquotesingle{}Or\textquotesingle{}}\NormalTok{: }\DecValTok{0}\NormalTok{, }\StringTok{\textquotesingle{}Ok\textquotesingle{}}\NormalTok{: }\DecValTok{0}\NormalTok{, }\StringTok{\textquotesingle{}Ode\textquotesingle{}}\NormalTok{: }\DecValTok{0}\NormalTok{, }\StringTok{\textquotesingle{}h\textquotesingle{}}\NormalTok{: }\FloatTok{0.67}\NormalTok{\}}
\NormalTok{param\_EdS }\OperatorTok{=}\NormalTok{ astro.param(cosmo}\OperatorTok{=}\NormalTok{cosmo\_dict)}

\NormalTok{cdist\_EdS }\OperatorTok{=}\NormalTok{ astro.comoving\_distance(param\_EdS, zarr)}
\NormalTok{pdist\_EdS }\OperatorTok{=}\NormalTok{ astro.proper\_distance(param\_EdS, zarr)}

\CommentTok{\# de Sitter universe}
\NormalTok{cosmo\_dict }\OperatorTok{=}\NormalTok{ \{}\StringTok{\textquotesingle{}Om\textquotesingle{}}\NormalTok{: }\FloatTok{0.0}\NormalTok{, }\StringTok{\textquotesingle{}Or\textquotesingle{}}\NormalTok{: }\DecValTok{0}\NormalTok{, }\StringTok{\textquotesingle{}Ok\textquotesingle{}}\NormalTok{: }\DecValTok{0}\NormalTok{, }\StringTok{\textquotesingle{}Ode\textquotesingle{}}\NormalTok{: }\FloatTok{1.0}\NormalTok{, }\StringTok{\textquotesingle{}h\textquotesingle{}}\NormalTok{: }\FloatTok{0.67}\NormalTok{\}}
\NormalTok{param\_dS }\OperatorTok{=}\NormalTok{ astro.param(cosmo}\OperatorTok{=}\NormalTok{cosmo\_dict)}

\NormalTok{cdist\_dS }\OperatorTok{=}\NormalTok{ astro.comoving\_distance(param\_dS, zarr)}
\NormalTok{pdist\_dS }\OperatorTok{=}\NormalTok{ astro.proper\_distance(param\_dS, zarr)}

\CommentTok{\# Benchmark model or Planck universe}
\NormalTok{cosmo\_dict }\OperatorTok{=}\NormalTok{ \{}\StringTok{\textquotesingle{}Om\textquotesingle{}}\NormalTok{: }\FloatTok{0.31}\NormalTok{, }\StringTok{\textquotesingle{}Or\textquotesingle{}}\NormalTok{: }\FloatTok{0.0}\NormalTok{, }\StringTok{\textquotesingle{}Ok\textquotesingle{}}\NormalTok{: }\FloatTok{0.0}\NormalTok{, }\StringTok{\textquotesingle{}Ode\textquotesingle{}}\NormalTok{: }\FloatTok{0.69}\NormalTok{, }\StringTok{\textquotesingle{}h\textquotesingle{}}\NormalTok{: }\FloatTok{0.67}\NormalTok{\}}
\NormalTok{param }\OperatorTok{=}\NormalTok{ astro.param(cosmo}\OperatorTok{=}\NormalTok{cosmo\_dict)}

\NormalTok{cdist }\OperatorTok{=}\NormalTok{ astro.comoving\_distance(param, zarr)}
\NormalTok{pdist }\OperatorTok{=}\NormalTok{ astro.proper\_distance(param, zarr)}

\CommentTok{\# Plots}
\NormalTok{fig, axs }\OperatorTok{=}\NormalTok{ plt.subplots(}\DecValTok{1}\NormalTok{,}\DecValTok{2}\NormalTok{,figsize}\OperatorTok{=}\NormalTok{(}\DecValTok{10}\NormalTok{,}\DecValTok{4}\NormalTok{))}
\NormalTok{axs[}\DecValTok{0}\NormalTok{].loglog(zarr, cdist\_dS, ls}\OperatorTok{=}\StringTok{\textquotesingle{}{-}.\textquotesingle{}}\NormalTok{, label}\OperatorTok{=}\StringTok{\textquotesingle{}de Sitter\textquotesingle{}}\NormalTok{)}
\NormalTok{axs[}\DecValTok{0}\NormalTok{].loglog(zarr, cdist, ls}\OperatorTok{=}\StringTok{\textquotesingle{}{-}\textquotesingle{}}\NormalTok{, label}\OperatorTok{=}\StringTok{\textquotesingle{}Planck\textquotesingle{}}\NormalTok{)}
\NormalTok{axs[}\DecValTok{0}\NormalTok{].loglog(zarr, cdist\_EdS, ls}\OperatorTok{=}\StringTok{\textquotesingle{}{-}{-}\textquotesingle{}}\NormalTok{, label}\OperatorTok{=}\StringTok{\textquotesingle{}Einstein{-}de Sitter\textquotesingle{}}\NormalTok{)}
\NormalTok{axs[}\DecValTok{0}\NormalTok{].legend(loc}\OperatorTok{=}\DecValTok{2}\NormalTok{, fontsize}\OperatorTok{=}\DecValTok{14}\NormalTok{)}
\NormalTok{axs[}\DecValTok{0}\NormalTok{].set\_xlabel(}\StringTok{\textquotesingle{}Redshift\textquotesingle{}}\NormalTok{, fontsize}\OperatorTok{=}\DecValTok{15}\NormalTok{)}
\NormalTok{axs[}\DecValTok{0}\NormalTok{].set\_ylabel(}\StringTok{\textquotesingle{}Comoving distance\textquotesingle{}}\NormalTok{, fontsize}\OperatorTok{=}\DecValTok{15}\NormalTok{)}
\NormalTok{axs[}\DecValTok{0}\NormalTok{].axis([}\FloatTok{0.01}\NormalTok{,}\DecValTok{10}\NormalTok{,}\DecValTok{40}\NormalTok{,}\FloatTok{5e4}\NormalTok{])}

\NormalTok{axs[}\DecValTok{1}\NormalTok{].loglog(zarr, pdist\_dS, ls}\OperatorTok{=}\StringTok{\textquotesingle{}{-}.\textquotesingle{}}\NormalTok{, label}\OperatorTok{=}\StringTok{\textquotesingle{}de Sitter\textquotesingle{}}\NormalTok{)}
\NormalTok{axs[}\DecValTok{1}\NormalTok{].loglog(zarr, pdist, ls}\OperatorTok{=}\StringTok{\textquotesingle{}{-}\textquotesingle{}}\NormalTok{, label}\OperatorTok{=}\StringTok{\textquotesingle{}Planck\textquotesingle{}}\NormalTok{)}
\NormalTok{axs[}\DecValTok{1}\NormalTok{].loglog(zarr, pdist\_EdS, ls}\OperatorTok{=}\StringTok{\textquotesingle{}{-}{-}\textquotesingle{}}\NormalTok{, label}\OperatorTok{=}\StringTok{\textquotesingle{}Einstein{-}de Sitter\textquotesingle{}}\NormalTok{)}
\NormalTok{axs[}\DecValTok{1}\NormalTok{].legend(loc}\OperatorTok{=}\DecValTok{2}\NormalTok{, fontsize}\OperatorTok{=}\DecValTok{14}\NormalTok{)}
\NormalTok{axs[}\DecValTok{1}\NormalTok{].set\_xlabel(}\StringTok{\textquotesingle{}Redshift\textquotesingle{}}\NormalTok{, fontsize}\OperatorTok{=}\DecValTok{15}\NormalTok{)}
\NormalTok{axs[}\DecValTok{1}\NormalTok{].set\_ylabel(}\StringTok{\textquotesingle{}Proper distance\textquotesingle{}}\NormalTok{, fontsize}\OperatorTok{=}\DecValTok{15}\NormalTok{)}
\NormalTok{axs[}\DecValTok{1}\NormalTok{].axis([}\FloatTok{0.01}\NormalTok{,}\DecValTok{10}\NormalTok{,}\DecValTok{40}\NormalTok{,}\FloatTok{5e4}\NormalTok{])}

\NormalTok{plt.tight\_layout()}
\NormalTok{plt.show()}
\end{Highlighting}
\end{Shaded}

\begin{figure}
\centering
\includegraphics{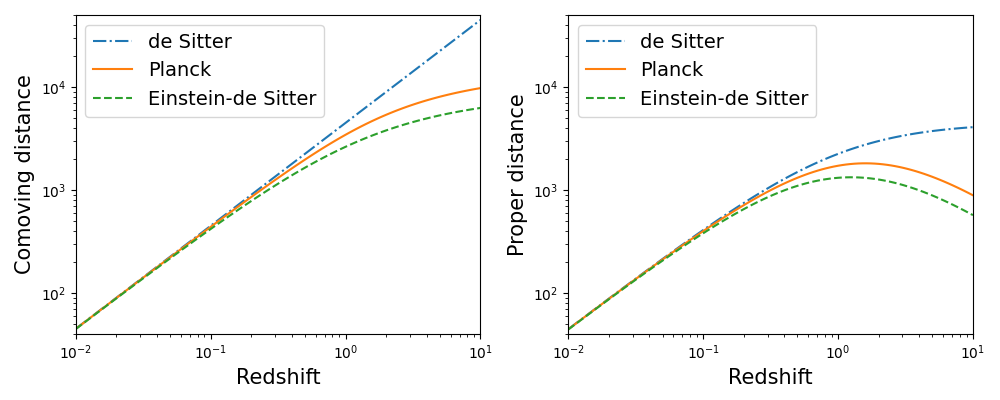}
\caption{Comoving (left) and proper (right) cosmological distances for
different redshifts plotted for three model universes.}
\label{fig:cosmodist}
\end{figure}

In Figure~\ref{fig:cosmodist}, we show the plot produced by the examples discussed above.
To understand the impact of the universe's expansion, we define two
important distances: comoving distance and proper distance. The comoving
distance represents how far apart two galaxies are, accounting for the
universe's expansion over time, and remains constant for objects moving
with the Hubble flow. In contrast, proper distance is what you would
measure if you could ``freeze'' the universe at a specific moment and
measure the physical separation between two objects, which changes as
the universe expands.

In the Einstein-de Sitter and Planck models, the proper distance
exhibits a turnover at high redshifts because, in the early universe, it
was much smaller and denser. As we look further back in time (to higher
redshifts), the proper distance between objects becomes shorter,
reflecting the universe's smaller scale. Thus, although proper distance
generally increases with redshift due to the universe's expansion, it
reaches a peak and then decreases at very high redshifts, corresponding
to a time when the universe was much more compact. In contrast, in the
de Sitter universe, we do not observe a strong turnover as the expansion
rate is consistently very high.

\section{Acknowledgements}\label{acknowledgements}

We acknowledge the higher education teaching and learning courses
offered by Stockholm University. Nordita is supported in part by
NordForsk.

\section*{References}\label{references}
\addcontentsline{toc}{section}{References}

\phantomsection\label{refs}

\end{document}